\documentclass[preprint,11pt]{aastex}
\usepackage{emulateapj5} 
\usepackage{psfig,epsfig}
\usepackage{apjfonts}

\def\spose#1{\hbox to 0pt{#1\hss}}
\def\lax{$\mathrel{\spose{\lower 3pt\hbox{$\mathchar"218$}}
     \raise 2.0pt\hbox{$\mathchar"13C$}}$}
\def\gax{$\mathrel{\spose{\lower 3pt\hbox{$\mathchar"218$}}
     \raise 2.0pt\hbox{$\mathchar"13E$}}$}
\def\x1{XMMU\,J013636.5$+$155036}

\def\xx2{XMMU\,J013638.6$+$154420}

\newcommand{\chandra}{{\it Chandra}}
\newcommand{\xmm}{{\it XMM-Newton}}
\newcommand{\lum}{\thinspace\hbox{$\hbox{erg}\thinspace\hbox{s}^{-1}$}}
\newcommand{\flux}{\thinspace\hbox{$\hbox{erg}\thinspace\hbox{cm}^{-2}
            \thinspace\hbox{s}^{-1}$}}

\slugcomment{submitted to ApJL}

\shorttitle{\xmm\ observations of M\,74}

\begin{document}

\title{\xmm\ observations of the spiral galaxy M\,74 (NGC 628)}

\author{Roberto Soria\altaffilmark{1} and Albert K.~H.~Kong\altaffilmark{2}}

\altaffiltext{1}{Mullard Space Science Laboratory,
  University College London, Holmbury St.~Mary, Surrey, RH5~6NT, UK; rs1@mssl.ucl.ac.uk}
\altaffiltext{2}{Harvard-Smithsonian Center for Astrophysics, 60
  Garden Street, Cambridge, MA 02138, USA; akong@cfa.harvard.edu}

\begin{abstract}

The face-on spiral galaxy M\,74 (NGC 628) was observed by \xmm\ on
2002 February 2.
In total, 21 sources are found in the inner 5\arcmin\ 
from the nucleus (after rejection of a few sources associated to foreground 
stars). Hardness ratios suggest that about half of them belong to the galaxy. 
The higher-luminosity end of the luminosity function is fitted 
by a power-law of slope $-0.8$. This can be interpreted as evidence 
of ongoing star formation, in analogy with the distributions 
found in disks of other late-type galaxies.
A comparison with previous \chandra\ observations 
reveals a new ultraluminous X-ray transient ($L_{\rm x} \approx 1.5 \times 10^{39}$ 
\lum\ in the 0.3--8 keV band) about 4\arcmin\ North of the nucleus. 
We find another transient black-hole candidate 
($L_{\rm x} \approx 5 \times 10^{38}$ \lum) about 5\arcmin\ North-West 
of the nucleus. The UV and X-ray counterparts of SN 2002ap
are also found in this \xmm\ observation. 
\end{abstract}



\section{Introduction}

M\,74 (NGC 628) is an SA(s)c galaxy, seen almost face-on (inclination 
angle $\approx 5^{\circ}$--$7^{\circ}$, Shostak and van der Kruit 1984) 
at a distance of 9.7 Mpc (Tully 1988). It was observed 
by \xmm\ on 2002 February 2, four days after the discovery of SN
2002ap (Nakano 2002) in the galaxy. In this Letter, we present the first
results (including the luminosity function of X-ray sources,
discovery of several luminous X-ray transients and the UV and X-ray
counterparts of SN 2002ap) from this observation. 

\section{Observations and Data Reductions}

%M\,74 was observed by \xmm\ on 2002 February 2. 
The \xmm\ instrument modes were 
full-frame, thin filter for the three EPIC cameras, 
and the UVW1 filter for the OM. 
The full time interval of the EPIC exposure is 34~ks. 
After rejecting intervals with a high background level, we considered 
a good time interval of 28~ks for our study. The exposure time for OM 
is 2.5~ks. The data were extracted and analyzed with the SAS version 5.2.
We also compared our \xmm\ observations with two \chandra\ observations
taken on 2001 June 19 and 2001 October 19. The exposure time of both
\chandra\ observations is about 47 ks. 
A full analysis of the \chandra\ data will be presented in a paper 
currently in preparation.

\section{Luminosity distribution of the X-ray sources}

Twenty-one sources are detected by \xmm/PN within 
5\arcmin\ from the galactic nucleus (not including a couple of sources 
coincident with foreground stars.)
Eighteen of those sources were also found by \chandra\  
in 2001 (one of them is in fact resolved into a close pair 
of sources by \chandra). However, three sources were not, 
and are likely to be transient. 
The completeness limit for the PN sources is 
$\sim 2 \times 10^{-3}$ cts s$^{-1}$ in the  
0.2--14 keV band. Assuming a power-law spectrum with 
$\Gamma = 1.7$ and column density $n_{\rm H} = 10^{21}$ cm$^{-2}$, 
this corresponds to a flux of $\sim 7 \times 10^{-15}$ \flux, 
and an emitted luminosity of $\sim 1 \times 10^{38}$ \lum.

Some of the 21 sources found inside the inner 5\arcmin\ may 
be background AGN, seen through the disk 
of the galaxy, and therefore highly extincted. 
We can estimate the relative contribution of 
background sources by using the number counts 
from the \chandra\ Deep Field North survey (Brandt et al.~2001) 
in the 2--8 keV band (where absorption is not important). 
We obtain that about half 
of the sources should be attributed to the background, 
at the completeness limit for the PN in the 2--8 keV band 
(flux of $\sim 5 \times 10^{-15}$ \flux).

We can estimate the minimum total column 
density of hydrogen that would be between us 
and a background AGN, and use the observed hardness 
ratios to decide which sources are more likely 
to belong to the galaxy.
By combining \ion{H}{1} (Shostak \& van der Kruit 1984) 
and CO (Adler \& Liszt 1989) radio observations, 
and using a standard CO-to-H$_{2}$ conversion factor (Bloemen 
et al.~1986), we infer that the total column 
density $N($\ion{H}{1}$+\rm{H}_2)$ goes from $\sim 3.5 \times 10^{21}$ 
cm$^{-2}$ in the inner 30\arcsec~to $\sim 1.5 \times 10^{21}$ cm$^{-2}$ 
at a distance of 4\arcmin, and $\sim 1 \times 10^{21}$ cm$^{-2}$ 
at 5\arcmin. To this, we need to add a foreground 
Galactic \ion{H}{1} column density $n_{\rm H, Gal} \approx 0.5 \times 10^{21}$ 
cm$^{-2}$ (Dickey \& Lockman 1990). Hence, we expect 
that background AGN would be seen through a total 
column density $n_{\rm H}$ \gax $1.5 \times 10^{21}$ cm$^{-2}$.

Defining as S and H the PN count rates in the (0.2--1.5) and (1.5--14) keV 
bands respectively, we obtain hardness ratios HR $=$ (H$-$S)/(H$+$S) 
for the detected sources. For a source with 
a power-law spectrum with $\Gamma = 1.7$, typical of an AGN, 
HR $= 0$ if $n_{\rm H} = 1.9 \times 10^{21}$ cm$^{-2}$, 
and HR $= -0.19$ if $n_{\rm H} = 1.0 \times 10^{21}$ cm$^{-2}$. 
We have therefore taken HR $< -0.2$ as an empirical criterion 
to identify sources very likely to belong to M\,74, 
while we cannot tell whether harder sources are 
background AGN or highly absorbed M\,74 objects. 

Nine of the 12 brightest sources 
detected with PN have HR $< -0.2$, hence we  
conclude that they are genuine M\,74 objects. 
Figure 1 shows the cumulative luminosity function 
log[N($>$S)]--log S for all the sources, and for the 
9 sources with a most likely galactic origin. The latter curve 
can be fitted with a simple power law of index $-0.8$. 
This is similar to the slope at the high-luminosity  
end of the luminosity function for the disk population 
in spiral galaxies such as M\,81 (Tennant et al.~2001) 
and M\,101 (Pence et al.~2001). 
We shall present a detailed discussion of the position, 
luminosity and hardness ratios of all \xmm\ and \chandra\ 
sources in a forthcoming paper. 

\section{Spectral Analysis of Individual Sources}

\subsection{A new Ultraluminous X-ray Transient}

The brightest source in the field of M\,74, located at
{R.A.~$=$ 1$^h$\,36$^m$\,36$^s$.5}, 
{Dec.~$=$ 15$^{\circ}$\,50\arcmin\,36\arcsec} (J2000), is 
one of the three sources found by \xmm\ but clearly below 
the detection limit ($\sim 10^{37}$ \lum) of both \chandra\ observations.
%Given that the luminosity of this source
%during our \xmm\ observations is \gax $10^{39}$ \lum, 
%a variability of two orders of magnitude 
%in a few months suggests the transient nature
%of \x1. 
%\subsection{Spectral Analysis}
Energy spectra were extracted with the SAS task {\sc xmmselect} and were
analyzed with XSPEC version 11.
%\footnote{Available at
%http://heasarc.gsfc.nasa.gov/docs/xanadu/xspec/index.html}. 
We extracted data in PN from a circle of $30$\arcsec\ centered on
the source, and the background from an annulus with inner and outer radii
of $45$\arcsec\ and $75$\arcsec, respectively. For the MOS images, in which the source is
near the chip boundary, we took two source-free
regions ($60$\arcsec\ radius each) near the source for the background.
In order to allow $\chi^2$
statistics to be used, all the spectra were binned such that the signal-to-noise 
ratio $\ge 4$.  
We fitted the PN and MOS2 data
simultaneously with several spectral models: power-law, 
black-body $+$ power-law, thermal bremsstrahlung, disk blackbody, 
and the bmc comptonisation model, in all cases with 
interstellar absorption. 
%The best-fit parameters determined by these
%fits are shown in Table 1.

All models provide statistically acceptable
fits to the data ($\chi^2$ \lax 1; Table 1).  The single power-law 
model yields a somewhat high column density compared 
to the values (foreground Galactic plus intrinsic) 
expected at that galactic radius.
In contrast, the disk blackbody model (Mitsuda et al.~1984), 
and the bmc model (Shrader \& Titarchuk 1999) provide lower values of $n_{\rm H}$. 
Physically, the comptonization model represents a two-component spectrum, 
with a soft blackbody component from the accretion disk 
and a power-law tail at higher energies, 
produced by inverse-Compton scattering of the soft photons 
by high-energy electrons.
The best-fitting bmc model is shown in Figure~2.  
Neither the bmc nor the blackbody $+$ power-law fits 
can simultaneously constrain the column density and the temperature 
of the thermal component.

The X-ray luminosity of \x1 is significantly above the Eddington
limit for a $1.4M_{\odot}$ accreting neutron star ($\sim 2\times10^{38}$
\lum). With the disk-blackbody model, the best-fitting parameters of 
$T_{in}=1.3 \pm 0.2$ keV and
$R_{in}\sqrt{\cos \theta}=47^{+12}_{-11}$ km imply 
a 0.3--8 keV luminosity of about
$1.4 \times 10^{39}$ \lum. The bmc model yields 
an emitted luminosity of $1.6 \times 10^{39}$ \lum.

\subsection{Other individual sources}

There is another bright X-ray transient 
%$\sim 2'$ west from \x1, 
at {R.A.~$=$ 1$^h$\,36$^m$\,27$^s$.2}, 
{Dec.~$=$ 15$^{\circ}$\,50\arcmin\,05\arcsec} (J2000); it was 
not seen in either \chandra\ observations. Its
spectrum can be well fitted with a simple power law
with photon index of $1.7^{+0.6}_{-0.3}$, 
$n_{\rm H}=(1.9^{+1.3}_{-1.4})\times 10^{21}$ cm$^{-2}$. The emitted 
0.3--8 keV luminosity is about
$5.2\times 10^{38}$ \lum, which makes it a likely
stellar-mass black hole (BH) candidate.
Compared to Galactic X-ray transients during outbursts, the spectrum
of this source is relatively hard, although the errors are too
large to constrain the exact spectral shape. Hard
X-ray transients are seen in other external galaxies
such as M\,31 (Trudolyubov, Borozdin, \& Priedhorsky 2001; Kong et
al. 2002) and M\,81 (Ghosh et al. 2001). 

A third possible transient, detected by \xmm\ but not \chandra, 
is located at {R.A.~$=$ 1$^h$\,36$^m$\,38$^s$.6}, 
{Dec.~$=$ 15$^{\circ}$\,44\arcmin\,20\arcsec} (J2000).
With a PN count rate of $(1.7 \pm 0.05) \times 10^{-3}$ cts s$^{-1}$, 
it is too faint for a spectral analysis. Assuming 
a power-law spectrum with $\Gamma = 1.7$, and 
column density $n_{\rm H} = 10^{21}$ cm$^{-2}$, 
its emitted luminosity is $\sim 8 \times 10^{37}$ \lum\
in the 0.2--14 keV band.
The nucleus of the galaxy is detected at a count rate of 
$(4.8 \pm 0.7) \times 10^{-3}$ cts s$^{-1}$, 
which corresponds to a luminosity of $\sim 2 \times 10^{38}$ \lum\ 
for the assumed spectral model.

Finally, the X-ray counterpart of SN 2002ap (Nakano 2002) 
is detected at a count rate of $(3.1 \pm 0.5) 
\times 10^{-3}$ cts s$^{-1}$. (Count rates in the two bands: 
S $=(1.4 \pm 0.4) \times 10^{-3}$ cts s$^{-1}$; 
H $=(2.0 \pm 0.5) \times 10^{-3}$ cts s$^{-1}$). 
This corresponds to $L \approx 6 \times 10^{37}$ \lum for thermal plasma 
emission at 0.5 keV, or $L \approx 10^{38}$ \lum 
for a thermal Bremsstrahlung at 2 keV.
SN 2002ap is also clearly detected in the OM image 
(Figure~3). We measure a flux $f_{\lambda,\rm{UV}} 
\approx 7.7 \times 10^{-15}$ \flux \AA$^{-1}$, 
at an effective wavelength of 291 nm, 
on MJD 52307.0153.

\section{Discussion}

\subsection{X-ray luminosity function in galactic disks}

The cumulative luminosity function is a useful tool 
for studying populations of X-ray sources in galaxies.
In some cases, it is a broken power-law, with a change in slope at 
luminosities varying between a few times $10^{37}$ and 
a few times $10^{38}$ \lum. In other galactic 
environments, the slope is constant 
from the detection limit to the high-luminosity end. 
A relation between the star-formation history of a galaxy 
and the evolution of its X-ray luminosity 
function has been proposed by Wu (2001) and Kilgard et
al. (2002). Assuming that the brightest X-ray sources also have the
shortest life, 
the luminosity function develops a break at its bright end 
in the absence of ongoing star formation. This is usually 
the case in elliptical galaxies and in the bulges 
of spiral galaxies. The break evolves towards lower 
luminosities over time. 
An unbroken power-law shape can instead be sustained 
if the X-ray source population is continuously replenished 
by star formation processes, as is the case for example 
in the arms of spiral galaxies.

In this scenario, the unbroken power-law luminosity function 
observed in the late-type spiral M\,74 is consistent 
with recent or ongoing star formation.
Previous optical and UV studies (Cornett et al.~1994) 
have shown evidence for recent star formation in this galaxy, 
probably not uniformly across the disk. 
They have also shown that the optical/UV colors of M\,74 
are intermediate between those of M\,33 (bluest, with 
more recent star formation) and of the M\,81 bulge (with 
an older stellar population).
A comparison between the luminosity functions of these 
three galaxies is left to a work in preparation.

\subsection{The nature of \x1}

Objects with an X-ray luminosity $\sim10^{39}$--$10^{41}$ \lum\ are often
called ultraluminous X-ray sources (ULXs; e.g., Makishima et al. 2000; 
Kaaret et al. 2001; Matsumoto et al. 2001).
The X-ray spectra of ULXs are well described by a disk blackbody model (see
however Strickland et al. 2001) with
$T_{in}\sim 1.0$--$1.8$ keV and $R_{in}\sqrt{\cos\theta} \sim 20$--$200$ km. Our
disk-blackbody fit to the \xmm\ spectrum of \x1 is 
in good agreement with these values.
Although some ULXs are highly variable on timescales of months or years
(e.g., Kubota et al. 2001; Strickland et al. 2001; La Parola et al. 2001),
they always remain at a high luminosity state (\gax\,$10^{39}$
\lum). However, \x1 is
undetectable in the two deep \chandra\ observations few months before our
\xmm\ observations, suggesting that the source was then below $10^{37}$ \lum. 
The only transient ULX found so far is CXOU\,J112015.8$+$133514 in NGC
3628 (Strickland et al. 2001), while  
\x1 is the only example of an off-center ($\sim 4'$ or $\sim 10$ kpc
from the galactic center) transient ULX.

The nature of ULXs is still a subject of debate. One common interpretation
is that they are powered by accretion onto an intermediate-mass BH 
($10^2$--$10^4$$M_{\odot}$; Colbert \& Mushotzky 1999). 
If the fitted spectral parameter $T_{\rm in}$ (color temperature of the 
disk-blackbody) corresponds to the effective temperature 
$T_{\rm eff}$ at the inner 
boundary of a Shakura-Sunyaev disk, values of $T_{\rm in}$ \gax 1 keV would be 
inconsistent with an intermediate-mass Schwarzschild 
BH. Either a rapidly-spinning BH (Makishima et al. 2000)
or non-standard disk models (e.g., a ``slim disk''; Watarai, 
Mizuno, \& Mineshige 2001) would be required 
instead. However, it is more generally $T_{\rm in} = fT_{\rm eff}$, 
where the ``hardening factor'' $f \approx 2.6$ in some Galactic 
microquasars (Ebisawa et al.~2001; Titarchuk \& Shrader 2002). 
Therefore, observed values $T_{\rm in} \approx 1$ keV 
do not rule out an intermediate-mass, non-rotating BH.
An alternative scenario for persistent ULXs is based 
on anisotropic emission from more common intermediate- or high-mass 
X-ray binaries. Emission beamed towards us can produce 
the high X-ray fluxes observed in those systems (King et al. 2001) 
without the need for an intermediate-mass BH. 
For a transient system such as \x1, further observations 
will tell us whether the brightening is due to a 
short-lived phase of super-Eddington accretion, or 
to a state transition like those observed 
in some Galactic microquasars (GRO\,J1655--40 and
GRS\,1915+105), which share similar spectral properties 
(e.g., Makishima et al. 2000).

We tried to ascertain whether the ULX can be associated with a supernova
remnant (SNR) or a photoionized \ion{H}{2} region, as found in some cases (Wang 2002). 
From archival H$\alpha$ images taken by the 1.0 m telescope at Mount
Laguna Observatory (Marcum et al. 2001), \x1 is $\sim 5$\arcsec\ North-West 
of an extended \ion{H}{2} region, and a point-like object is within 
the X-ray error circle (see
Figure 4). The point-like source appears more or less
the same in both continuum-subtracted H$\alpha$ images taken in 1991 and 1995. 
Optical spectroscopy of the extended \ion{H}{2} region 
yields a ratio [\ion{S}{2}]/H$\alpha$ $\approx 0.18$ (van Zee et al. 1998), 
significantly lower than the typical value ($>0.4$) measured in 
SNR (e.g., Blair, Kirshner, \& Chevalier 1981). Therefore, the 
extended H$\alpha$ feature near \x1 is likely a photoionized 
\ion{H}{2} region associated with star formation.

We inspected the UV image taken with OM, simultaneously with the X-ray images. 
There is no obvious object within the X-ray error circle, 
to a limiting specific flux of $\sim 2 \times 10^{-17}$ \flux 
\AA$^{-1}$, at an effective wavelength of 291 nm. As a comparison,
the specific flux from a single B0\,V star in M\,74 would 
be $\sim 10$ times fainter at the same 
effective wavelength.  
If the point-like source in the H$\alpha$ image at the position 
of \x1 corresponds to an OB star cluster, the lack of UV detection 
may be due to high extinction, typical 
of a very young star-forming cluster (Marcum et al. 2001).
The larger \ion{H}{2} region South-East 
of the transient is instead well visible in
the OM image (Figure 3).

The association with a young stellar population 
suggests that \x1 is related to a high-mass X-ray binary.
In addition, its strong variability ($> 100\times$) 
indicates that it is an accreting compact object 
rather than a SNR. Further observations at different 
wavelengths will be necessary to shed light on the nature and 
timing properties of the source.

\section{Conclusions}

We have studied the spiral galaxy M\,74 with \xmm. 
The luminosity function above $\sim 10^{38}$\lum\ 
is well fitted by a single power law of slope $-0.8$, 
similar to the values found for the source population 
in the disks of other spiral galaxies. We interpret it  
as evidence of ongoing star formation in the disk. 
The brightest source, located $\sim 4$\arcmin\ from 
the nucleus, is a transient, undetected by \chandra\ 
a few months earlier. It has a luminosity 
$L_{\rm x} \approx 1.5 \times 10^{39}$ 
\lum in the 0.3--8 keV band, and spectral features 
similar to those observed in ULXs in other galaxies.
The source appears to be associated with 
a photoionized \ion{H}{2} region. Another bright 
transient BH ($L_{\rm x} \sim 5 \times 10^{38}$ 
\lum) is found at $\sim 5$\arcmin\ from the nucleus. 
The nuclear source itself is weaker ($L_{\rm x} \approx 2 \times 10^{38}$ 
\lum). We have also measured the luminosity 
of the X-ray and UV counterparts of SN 2002ap, 
4 days after its optical discovery.

\begin{acknowledgements}
A.K.H.K. acknowledges the support of NASA LTSA Grants NAG5-10889 and
NAG5-10705.
\end{acknowledgements}

\begin{deluxetable}{lcccccc}
\tablewidth{0pt}
\tablecaption{Best-fitting Spectral Parameters for \x1}
\tablehead{Model & $n_{\rm H}$ & $\Gamma$ & $kT$ &
$R_{in}\sqrt{\cos \theta}$& $\chi^2_{\nu}$/dof & Flux\tablenotemark{a} \\
       &       ($10^{21}$ cm$^{-2}$) &  & (keV) & (km) &}
\startdata
power-law & $2.4^{+0.4}_{-0.5}$ & $2.0^{+0.2}_{-0.1}$ & & &0.85/56 & 1.95\\[5pt]
%Blackbody & 0\tablenotemark{c} & & $0.53\pm0.04$ & & 1.74/53& 0.75\\[5pt]
bremsstrahlung & $1.4^{+0.4}_{-0.3}$ && $4.6^{+1.4}_{-1.0}$ & &0.91/56 &1.53\\[5pt]
disk blackbody & $0.4^{+0.3}_{-0.4}$ & & $1.3^{+0.2}_{-0.2}$\tablenotemark{b} & $47^{+12}_{-11}$&
1.14/56& 1.26 \\[5pt]
bmc & $1.3^{+1.4}_{-0.3}$ &$1.9^{+0.2}_{-0.4}$& $0.23^{+0.09}_{-0.23}$ 
& &0.86/54 &1.49\\[5pt]
bb$+$power-law & $2.3^{+1.9}_{-0.6}$ &$1.9^{+0.3}_{-0.3}$& $0.22^{+1.53}_{-0.22}$ 
& &0.86/54 &1.89\\[5pt]
\enddata
\tablecomments{All quoted uncertainties are 90\% confidence.}
\tablenotetext{a}{Unabsorbed flux in 0.3--8 keV ($10^{-13}$ ergs
cm$^{-2}$ s$^{-1}$).}
\tablenotetext{b}{Color temperature $T_{\rm in}$.}
%\tablenotetext{c}{The value of $N_H$ hit the minimum of 0 allowed by XSPEC.}

\end{deluxetable}

%\end{document}

\begin{figure}[b]
%\begin{center}
%\plotone{lumdistr3.ps}
\epsfig{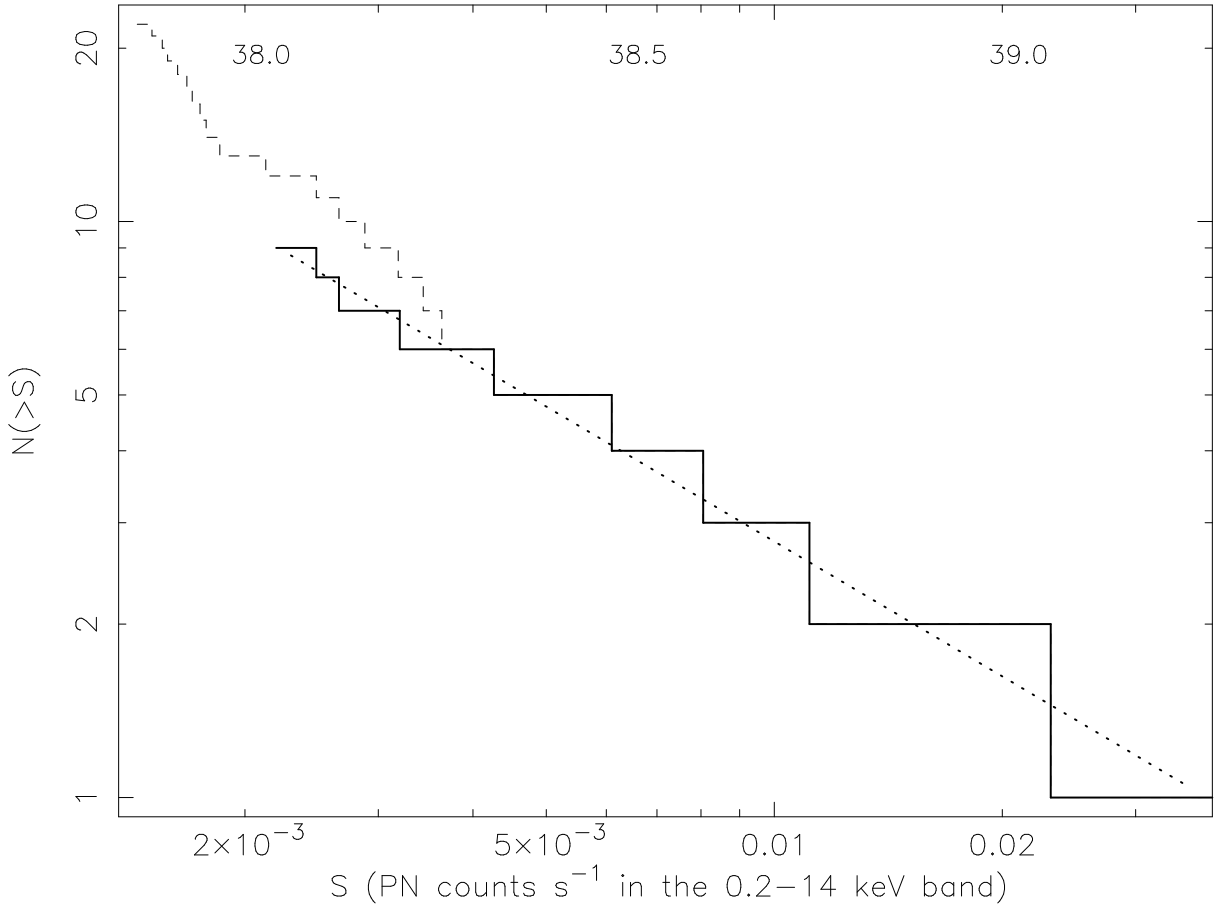}
%\end{center}
\caption{The log [N($>$S)]--log S curve of the 
brightest sources in M\,74 is well fitted with a power law 
of index $-0.8$ (dotted line). 
%This is analogous to the higher-luminosity 
%end of the luminosity functions in the disk 
%of other spiral galaxies with ongoing star formation. 
The thick solid line includes only relatively unabsorbed sources 
(HR $<-0.2$), likely to belong to the galaxy. The dashed line 
includes all the 21 sources found by \xmm\ within a 5\arcmin\ radius.
At fainter fluxes, most of the sources may be background AGN.
For the conversion between \xmm/PN counts and unabsorbed luminosity, 
we have assumed power-law spectra with $\Gamma = 1.7$, 
$n_{\rm H} = 10^{21}$ cm$^{-2}$, $d = 9.7$ Mpc.}
\end{figure}

%\end{document}

\begin{figure}[t]
\begin{center}
{\rotatebox{-90}{\psfig{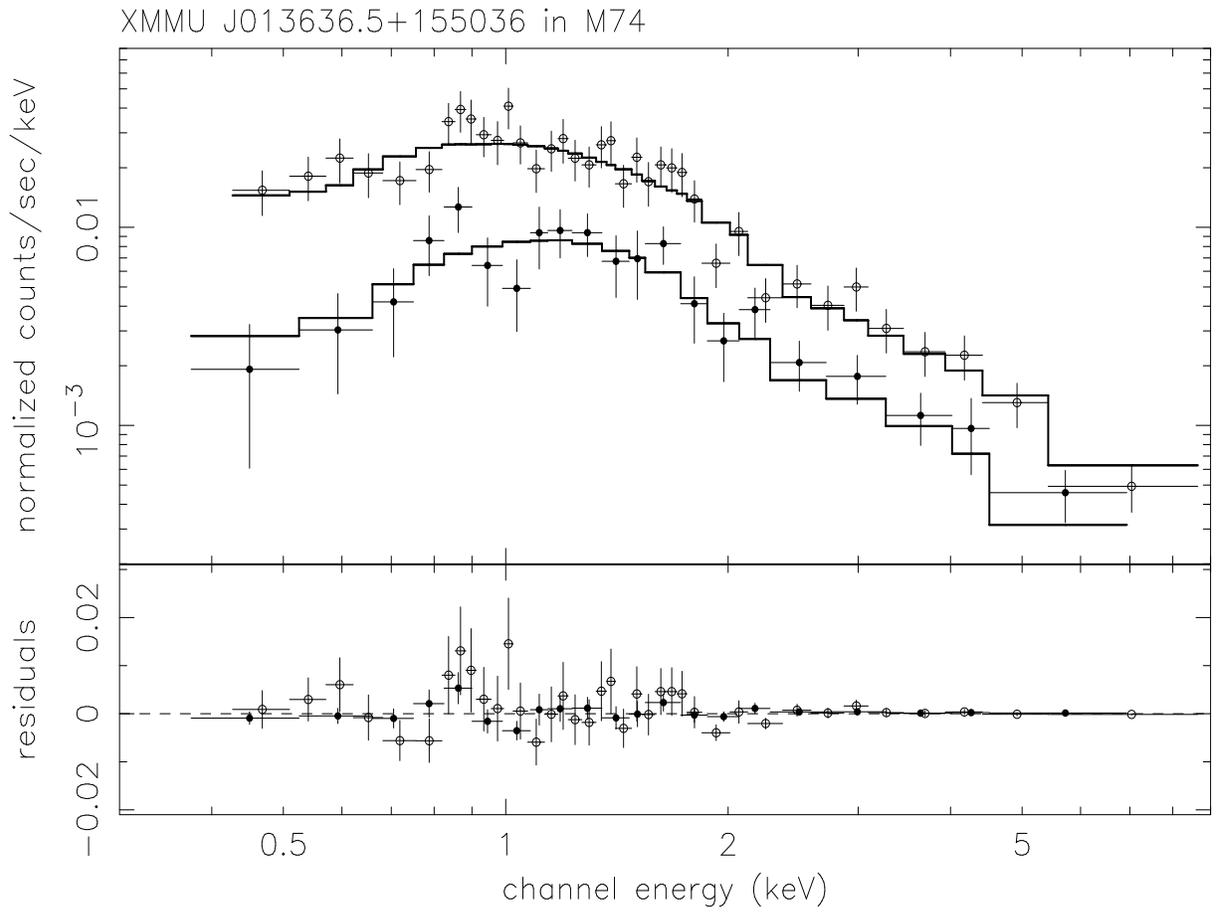}}}
\end{center}
\caption{\xmm/PN (open circles) and MOS2 (filled circles) spectra of
the ultraluminous X-ray transient \x1, fitted with an absorbed bmc model 
(parameters in Table 1).}
\end{figure}

\begin{figure}[t]
\vspace{-1cm}
\begin{center}
\epsfig{figure=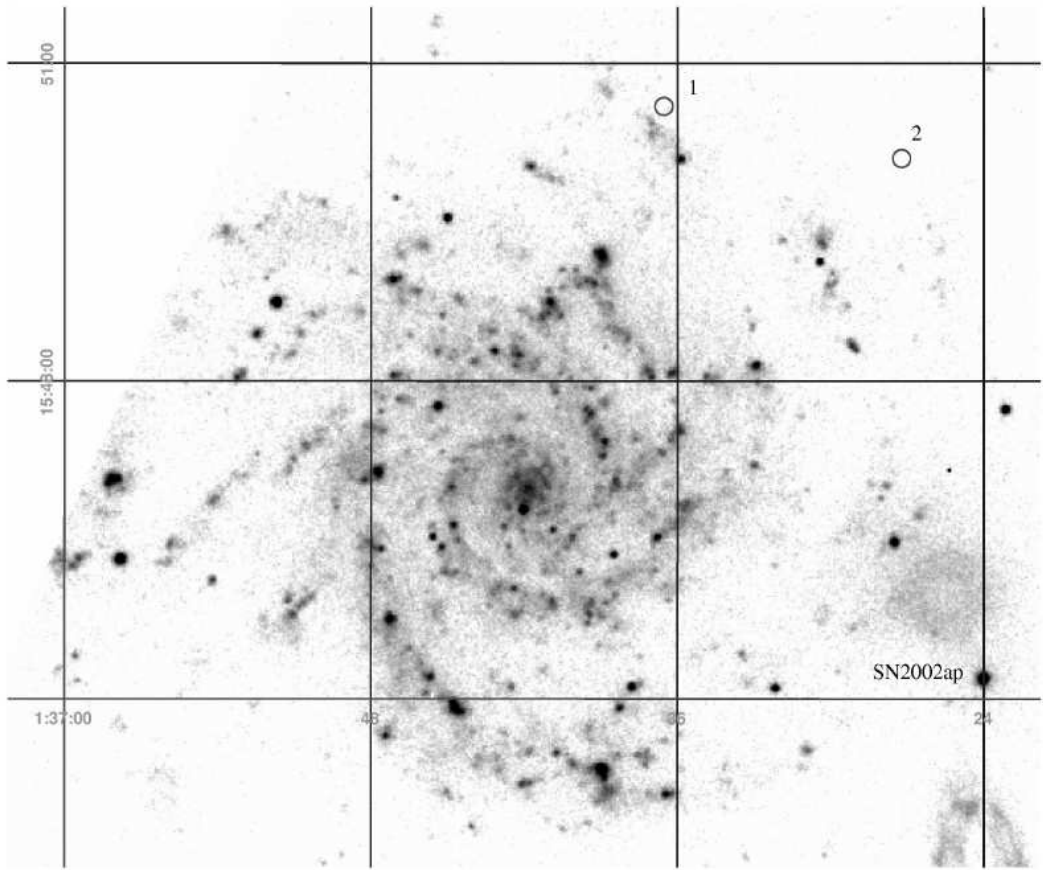,width=18cm}
\end{center}
\caption{
SN 2002ap is clearly visible at the bottom right 
of the \xmm/OM image, taken in the UVW1 filter (North is up, 
East is left).
%, at {R.A.~(2000) $=$ 1$^h$\,36$^m$\,24$^s$}, 
%{Dec.~(2000) $=$ 15$^{\circ}$\,45\arcmin\,13\arcsec}. 
The position of the two brightest X-ray transients 
is indicated with the circles (radius $= 5$\arcsec) 
near the top of the image: 1 $=$ \x1, 2 $=$ \xx2. 
(The two extended ring features at the bottom right 
of the image are due to reflected 
light onto the OM detector).
}
\end{figure}

\begin{figure}[t]
\begin{center}
\psfig{file=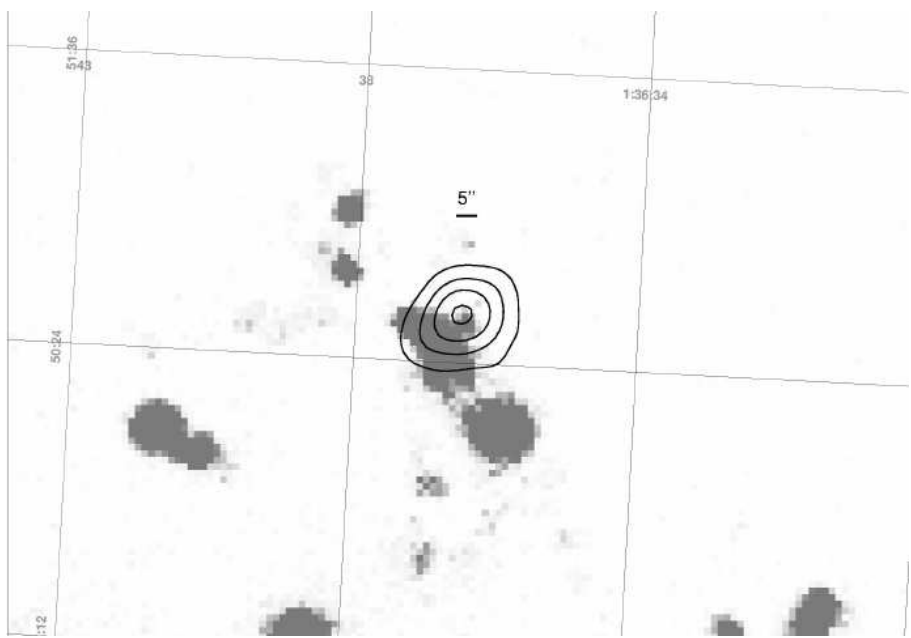,width=12cm}
\end{center}
\caption{
\xmm\ PN intensity contours (at 10\%, 25\%, 50\%
and 90\% of the maximum pixel value) of \x1 overlaid on an
archival continuum-subtracted H$\alpha$ image taken from 
the Mount Laguna Observatory 1.0-m telescope in 1995.}
\end{figure}

%\end{document}

\end{document}